\documentclass[rnote]{aa} 
\usepackage{graphicx}
\usepackage{txfonts}
\begin{document}
\hyphenation {at-tempt spots rea-ches Ge-min-ga Mi-la-gro km ef-fects sce-na-rio 
mo-del ex-plo-sions reaches value dif-fu-sion res-pect ca-ses vo-lume ge-ne-ra-ted}
   \title{The Milagro anticenter hot spots: \\
          cosmic rays from the Geminga supernova ?}

   \author{M. Salvati
          \inst{1}
          \and
          B. Sacco\inst{2}
          }

   \offprints{M. Salvati}

   \institute{INAF--Osservatorio Astrofisico di Arcetri \\
              Largo Enrico Fermi 5, I--50125 Firenze, Italy\\
              \email{salvati@arcetri.astro.it}
         \and
             INAF--Istituto di Astrofisica Spaziale e Fisica Cosmica\\
             Via Ugo La Malfa 153, I--90146 Palermo, Italy\\
             \email{bruno.sacco@iasf-palermo.inaf.it}
             }

   \date{Received 15 February 2008 / Accepted 10 April 2008}

 
  \abstract
   {The Milagro experiment has announced the discovery of an excess flux 
    of TeV cosmic rays from the general direction of the heliotail,
    also close to the Galactic anticenter.}
   {We investigate the hypothesis
    that the excess cosmic rays were produced in the SN explosion that
    gave birth to the Geminga pulsar.}
   {The assumptions underlying our proposed scenario are that the Geminga
    supernova occurred about 3.4~10$^5$ years ago (as indicated by the spin down 
    timescale), that a burst of cosmic rays was injected with total energy $\sim 10^{49}$
    erg (i.e., about 1\% of a typical SN output), and that the Geminga pulsar was born
    with a positive radial velocity of 100--200~km~s$^{-1}$.}
   {We find that our hypothesis is consistent with the available information. 
    In a first variant (likely oversimplified), the cosmic rays have diffused 
    according to the Bohm prescription (i.e., with a diffusion
    coefficient on the order of $c\times r_L$, with $c$ the speed of light
    and $r_L$ the Larmor radius). An alternative scheme assumes that diffusion 
    only occurred initially, and the final propagation to the Sun was a free 
    streaming in a diverging magnetic field.}
   {If the observed cosmic ray excess does indeed arise from the Geminga SN
    explosion, the long--sought ``smoking gun" connecting cosmic rays with 
    supernovae would finally be at hand. It could be said that, while looking
    for the ``smoking gun", we were hit by the bullets themselves.}


   \authorrunning{M. Salvati and B. Sacco}
   \titlerunning{Cosmic rays from the Geminga supernova?}

   \maketitle
%

\section{Introduction}

In a recent paper (\cite{milagro}), the Milagro collaboration reports the
detection of an excess of cosmic rays from the general direction of
the heliotail, also close to the Galactic anticenter. The excess is 
diffuse but confined (in the following we assume a solid 
angle of 0.3 steradians to account for both hot spots A and B), 
is composed of hadrons (photons and electrons
are excluded to a highly significant level) and has a harder spectrum
than the general cosmic ray background up to about 10 TeV.

The authors discuss some possible explanations and conclude that none of
them is viable. In particular, because of the positional coincidence of the excess
with the heliotail, they consider a local origin of the phenomenon,
but discard it on the ground that several--TeV particles could not be
easily generated or confined by the heliosphere. 

In this Note, we revisit the hypothesis of a heliospheric origin
of the hot spots and provide an additional quantitative argument
against it. Then we point out that the closest plausible extra solar
source is the supernova that produced the Geminga pulsar, and show
that there is a region in the parameter space where this alternative 
hypothesis is valid.

\section{The heliospheric scenario}

All known effects of the heliosphere on the cosmic rays (solar
modulation, anomalous cosmic rays) are detected at energies of about 
1~GeV per nucleon (e.g., \cite{helio}), much lower than the 
several TeV observed by Milagro. On top
of this, one can give a direct counter argument based on the energy budget.
In the following, we take data about the cosmic rays from Longair (1981) 
and data about the Very Local InterStellar Medium (VLISM) from Axford and
Suess (1994).

From the Milagro paper (their Fig. ~4) one deduces that the excess flux 
at 10~TeV --measured as a fraction of the background cosmic ray flux--
amounts to $15~10^{-4}$ in region A and $6~10^{-4}$ in region B. These 
values refer to the cores of the two regions, which have a relatively
small angular extent (for instance, the core of region A is only 0.02 
steradians). To account for the lower level excess that is visible 
around the cores, we add two times the core counts in a solid angle of 0.3 
steradians. Under these assumptions the average fractional excess is 
$5~10^{-4}$, and the excess flux turns out to be

\begin{equation}
\Phi\sim 5~10^{-4} \times 0.3 \times 6.7~10^{-6}\sim 1.0~10^{-9}~\rm erg~cm^{-2}~s^{-1}.
\end{equation}

\noindent The acceleration region must be at least as large as one gyration radius, 
which for a 10~TeV proton in a 1~$\mu$G magnetic field is $r_L = 0.01~\rm parsec$. 
Given the angular extent of the excess, the distance to the acceleration region
must also be at least as much as $r_L~$, i.e., at least 20~times the distance
to the heliopause. In the heliospheric hypothesis, the acceleration region is
powered by the converging flows of the solar wake, and, at the inferred distance,
the convergence angle must be $\sim1/20$~radians. Finally, the power dissipated 
in the acceleration region (for a local velocity $v$ of 25~km~s$^{-1}$ and a local 
density $n$ of 0.1~cm$^{-3}$), is approximately

\begin{equation}
P \sim r_L^2 n m_H \left(\frac{v}{20}\right)^3 \sim 3.1~10^{23}~\rm erg~s^{-1}. 
\end{equation}

\noindent This corresponds to a maximum flux at Earth equal to

\begin{equation}
\Phi \sim \frac{P}{4\pi r_L^2} \sim 2.6~10^{-11}~\rm erg~cm^{-2}~s^{-1},
\end{equation}

\noindent which is much too low with respect to the measured one.

\section{The supernova scenario, part one}

The basis of the supernova scenario is twofold. First, the Milagro excess 
flux comes from the right direction in the sky: hot spots A and B are 
about 50 degrees apart, and nicely encompass Geminga (e.g., \cite{geminga}). 
The pulsar has a non--negligible proper motion (125~km~s$^{-1}$ at a distance
of 155 parsec) so that its position at birth was different from the present one, 
the more so if the birthplace was close to the Sun. For the cases of interest, 
the displacement is around 20--30 degrees towards the south of 
region A. Such angular distances do not seem implausible in view of the effect
of the magnetic field on the arrival direction. 
Second, a 10~TeV proton diffusing in a 1~$\mu$G magnetic field 
in the Bohm regime (i.e., $ D \sim c \times r_L $) reaches an $e$--folding distance of 
65 parsec in the time elapsed since the Geminga supernova explosion 
($t_{exp} = 3.4~10^{5}$~yr, if the pulsar spin down age is adopted):

\begin{equation}
R = \sqrt{4 D t_{exp}} = \sqrt{4 c r_L t_{exp}} = \rm 65~pc.
\end{equation}

The diffusion of cosmic rays in the Galaxy (e.g., \cite{blasi}) is 
described with two diffusion coefficients,
$D_{\parallel}$ and $D_{\perp}$, parallel and orthogonal to the magnetic
field, respectively. The coefficients may be written as

\begin{equation}
D_{\parallel} \sim c \times \lambda, \quad D_{\perp} \sim c \times
\frac{r_L^2}{\lambda}
\end{equation}

\noindent where $\lambda$ is the mean free path along the magnetic field.
Usually $\lambda$ is taken much greater than $r_L$, so that $D_{\parallel}$
is much larger than $D_{\perp}$ and the diffusion is strongly anisotropic. 
However, there are circumstances where the two become equal to each other 
and the diffusion is isotropic in the Bohm regime: this happens  
in the limiting case of a very chaotic magnetic field, with $({\delta B}/{B}) 
\sim 1$ over distances $\sim r_L$, so that $\lambda \sim r_L$. 
It might be argued that the explosion of the Geminga supernova in the
relatively recent past has stirred the local interstellar magnetic
field up to the required level of chaos. We pursue this hypothesis
as a zeroth order assumption, useful for simple calculations, and a more
articulated scenario will be presented in the next section.

The present distance to Geminga is estimated to be $155^{+60}_{-35}\rm~pc$,
so we must assume a non negligible velocity of the pulsar in the
positive radial direction, equal to at least 160~km~s$^{-1}$. On
the one hand, such a radial velocity is discordant with
the morphology of the Geminga trail (\cite{bow}), which suggests a velocity
vector within 30 degrees from the plane of the sky. Based on this result,
Pellizza et al. (2005) put a lower limit of 90 pc on the distance from
the Sun at which the supernova explosion might have occurred. On the other
hand, a radial velocity as high as assumed here does not seem implausible with 
respect to the measured transverse velocity; moreover, an even higher value
has been suggested (\cite{neil}) in an attempt to relate the Geminga
supernova to the formation of the Local Bubble. In any case, the scenario
proposed in the next section can accomodate a distance at the lower limit
of Pellizza et al. (2005).

The density distribution of particles diffusing with a constant 
coefficient in a 3--dimensional region is

\begin{equation}
n(r,t) = N \frac{2}{3\sqrt{\pi}} \frac{e^{-\left(\frac{r^2}{4Dt}\right)}}
{\frac{4\pi(4Dt)^{3/2}}{3}}
\end{equation}

\noindent where $N$ is the total number injected in a small volume at 
$r=0, t=0$. If $\epsilon$ is the energy of the particles, the net 
flux at radius $r$ and time $t$ is 

\begin{equation}
\Phi = -\epsilon D \frac{\partial n(r,t)}{\partial r} =  
\epsilon n(r,t) \frac{r}{2t}\, .
\end{equation}

\noindent If we set $t=t_{exp}$, $\epsilon = 10$~TeV, and $r=R$ and require agreement
with the right hand side of Eq.(1), we deduce $n$ from Eq.(7) and $N$ from
Eq.(6). Finally, we assume that the excess cosmic rays have the same spectrum
as the background cosmic rays, and obtain the following estimate for the
cosmic ray output $E$ of the Geminga supernova

\begin{equation}
n(R,t_{exp}; \epsilon) = 6.7~10^{-18}~\rm cm^{-3}, \quad \it N(\epsilon) \rm = 1.7~10^{45}
\end{equation}
\begin{equation}
\it E \rm = 1.5~10^{49}~erg.
\end{equation}

\noindent This estimate is perfectly in line with the commonly required 
efficiency (about 1\%) with which a supernova energy output must be
channeled into cosmic rays if indeed supernovae are to maintain the
Galactic cosmic ray reservoir. 

The irregular distribution of the excess flux, and especially the
presence of two disjoint hot spots, is perhaps a consequence of 
large--scale irregularities in the background medium and background
magnetic field: the shape of the diffusing cloud must be 
much more complex than a perfect sphere.

A final comment is in order about assuming a spectrum of the excess 
cosmic rays similar to the one of the background cosmic rays, while
the Milagro data indicate a much flatter slope (1.5 versus 2.6) and
a cutoff above several TeV. The qualitative explanation that we 
propose has to do with the dependence of $r_L$ and $D$ on the particle energy.
If at about 10~TeV the $e$--folding point of the diffusing cloud profile
has reached the Sun, at much higher (lower) energies the $e$--folding
point is much beyond (before) the Sun position. The ratio of the excess
flux at a generic energy to the one at the fiducial energy (10~TeV)
can be expressed as a function of the ratio of the relevant diffusion
coefficients $D$ and $D_{10}$

\begin{equation}
\left(\frac{D_{10}}{D}\right)^{3/2} e^{\left(1-\frac{D_{10}}{D}\right)}\, ,
\end{equation}

\noindent so that in both limits the excess flux is diminished with respect 
to the fiducial case.

\section{The supernova scenario, part two}

Drury and Aharonian (2008) have raised two important objections to the
scheme presented above. First, a diffusing cloud of cosmic rays would
produce a very wide signal in the sky, instead of the relatively narrow
hot spots detected by Milagro. Second, while diffusion in the Bohm regime
is thought to occur in peculiar regions, it cannot be the general process
governing the 
propagation of cosmic rays across large distances. They suggest that
the hot spots might be due to excess cosmic rays streaming almost freely 
from a magnetic nozzle along a diverging field; the source of excess
cosmic rays should be relatively nearby (100 pc or less), located at 
the nozzle or behind it, and its energy content should be a fraction of a 
supernova output.

We note that the cosmic rays cannot stream freely all the way from the
source to the Sun, because the propagation time would be too short, less than
a thousand years, and such a young supernova remnant could not remain
unnoticed. Moreover, the scenario of Drury and Aharonian (2008) does not 
provide an explanation for the peculiar spectral shape of the Milagro 
signal. One must assume that propagation occurs by diffusion from the source 
to the magnetic nozzle, over a time long enough to allow the dissipation 
of the SNR. This initial part of the propagation process is very similar 
to what we discussed in the previous section. In particular, 
diffusion would again act as a ``passband'' filter in energy, producing
a hard spectrum with a high energy cutoff, analogous to the observed one.
In the new scheme we can relax the Bohm assumption $\lambda \sim r_L$,
$\rm D_{\parallel} \sim D_{\perp}$, and can accomodate a wider range of 
values for the distance of the supernova: indeed, at variance with the 
previous scheme, now the diffusing cloud of cosmic rays does not need to propagate 
from the supernova to the Sun, but only from the supernova to the magnetic nozzle.
More precisely, diffusion is only needed from the rim of the supernova remnant
to the magnetic nozzle, and across--field diffusion (which is the slowest
process of all) is needed only from the rim of the SNR to the first
``useful'' magnetic line (see Fig.~1).

We conclude by arguing that it is unlikely that the supernova responsible
for the excess cosmic rays was not the same that produced Geminga.
In a cone with vertex on the Sun, axis in the direction of Geminga, 
height 150, and base radius 50 pc, we expect less than 0.01 supernovae
in 3.4~10$^5$ years for a Galactic rate of 0.01~yr$^{-1}$.

   \begin{figure}[h!]
   \resizebox{\hsize}{!}{\includegraphics{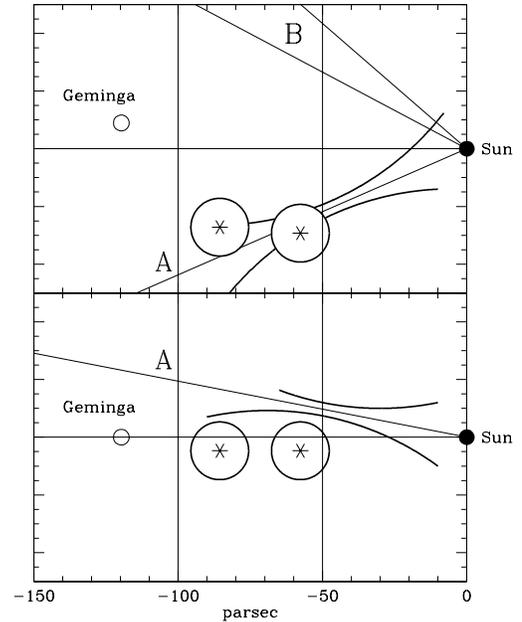}}
   \caption{Projection of the anticenter region on the meridian
     plane at Galactic longitude 195$^{\circ}$ (upper panel) and on the
     Galactic plane (lower panel). See text for details.}
   \end{figure}

In Fig.~1 we schematize the proposed geometry. The upper panel is the
projection of the anticenter region on the plane that, being orthogonal 
to the Galactic plane, contains the present positions 
of the Sun and Geminga (the latter at the ``close'' value of 120~pc). 
The lower panel is the projection of the same region on the Galactic plane 
itself. The directions of the Milagro hot spots are indicated, with the 
exception of hot spot B, which is not drawn in the lower panel since it is
very wide in Galactic longitude and would cover all other elements.
The two stars mark two possible positions of the supernova explosion, placed
at distances of 90 and 65 pc from the present position of the Sun, 
respectively. The circles have radii of 10 pc, and indicate the volume
occupied by fully developed supernova remnants. A possible magnetic nozzle 
is sketched with heavy lines: one sees that the cosmic 
rays, after leaving the remnant, need only diffuse across the field for 
very few parsecs before catching the right field line and propagating
to the Sun. 

Region B is 50 degrees away from Region A and could have a different origin.
However, this hypothesis would entail several additional {\it ad hoc}
assumptions and would, in any case, conflict with the low probability of
multiple nearby supernova explosions. It is still more economical to
attribute both regions to the same source. Then the large angular separation 
should be ascribed to large irregularities of the magnetic field in the Solar 
vicinity: some of the field lines diverging from the ``primary" magnetic nozzle 
intersect the line of sight to region B, as shown in the upper panel, and could
fuel a ``secondary" magnetic nozzle in that position. Admittedly, this is
another epicycle in the model, but it would have to be invoked in any case,
even if region B were fueled by a source independent of region A.

\section{Conclusions}

Our discussion does not {\it prove} that the Milagro anticenter hot 
spots are a transient relic of the explosion of the Geminga supernova. 
However, it does provide a consistent framework for 
an experimental result that would otherwise remain unexplained.

If our proposed picture were indeed true, one would conclude that
after all supernovae do produce cosmic rays with an efficiency of 
about 1\%. We could see this transient relic only because the supernova 
that gave birth to the Geminga pulsar exploded very nearby, and not very
long ago. Because of such lucky circumstances, the long sought 
``smoking gun" connecting cosmic rays with supernovae would finally 
be at hand. Indeed, it could be said that while looking for the
``smoking gun" we were hit by the bullets themselves.

\begin{acknowledgements}
We would like to thank the friends and colleagues who read our manuscript,
almost equally divided between strongly supportive and
strongly skeptical. They are Pasquale Blasi, Gabriele Ghisellini,
Franco Pacini, Malcolm Walmsley and Lodewijk Woltjer, but we
will not disclose who said what.

We also thank the referee, Patrizia Caraveo, for her patience, accuracy, and
open--mindedness in dealing with our wild ideas.

\end{acknowledgements}

{}

\end{document}